\documentclass[twocolumn,trackchanges]{aastex62}

\usepackage{amsmath}
\usepackage{graphicx}
\newcommand{\unitstyle}[1]{\ensuremath{\mathrm{#1}}}

\newcommand{\Msun}{\ensuremath{\unitstyle{M}_\odot}}

\newcommand{\code}[1]{\texttt{#1}}
\newcommand{\mesa}{\code{MESA}}
\newcommand{\MESA}{\mesa}

\input{vectors.tex}
\newlength{\apjcolwidth}
\newlength{\figwidth}
\newlength{\doublewide}
\begin{document}
\title{On Stellar Evolution In A Neutrino Hertzsprung-Russell Diagram}

\shorttitle{Neutrino HR Diagrams}
\shortauthors{Farag et al.} 

\author[0000-0002-5794-4286]{Ebraheem Farag}
\affiliation{School of Earth and Space Exploration, Arizona State University, Tempe, AZ 85287, USA}

\author[0000-0002-0474-159X]{F.X.~Timmes}
\affiliation{School of Earth and Space Exploration, Arizona State University, Tempe, AZ 85287, USA}
\affiliation{Joint Institute for Nuclear Astrophysics - Center for the Evolution of the Elements, USA}

\author[0000-0002-5107-8639]{Morgan Taylor}
\affiliation{School of Earth and Space Exploration, Arizona State University, Tempe, AZ 85287, USA}

\author[0000-0002-2154-4782]{Kelly M.\ Patton}
\affiliation{Department of Physics and Astronomy, Colby College, Waterville, ME 04961, USA}

\author[0000-0003-3441-7624]{R. Farmer}
\affiliation{Anton Pannenkoek Institute for Astronomy and GRAPPA, University of Amsterdam, NL-1090 GE Amsterdam, The Netherlands}

\correspondingauthor{Ebraheem Farag}
\email{ekfarag@asu.edu}

\begin{abstract}
%
% 250 word maximum
% this version is 185 words
%
\noindent

We explore the evolution of a select grid of solar metallicity stellar
models from their pre-main sequence phase to near their final fates in
a neutrino Hertzsprung-Russell diagram, where the neutrino luminosity
replaces the traditional photon luminosity.  Using a calibrated
\MESA\ solar model for the solar neutrino luminosity 
($L_{\nu,\odot}$ = 0.02398 $\cdot$ $L_{\gamma,\odot}$ = 9.1795 $\times$ 10$^{31}$ erg s$^{-1}$) as a normalization, 
we identify $\simeq$\,0.3 MeV electron neutrino emission from helium burning during the
helium flash 
(peak $L_{\nu} / L_{\nu,\odot} \simeq$ 10$^4$, flux $\Phi_{\nu, {\rm He \ flash}} \simeq$ 170 (10 pc/$d$)$^{2}$ cm$^{-2}$ s$^{-1}$
for a star located at a distance of $d$ parsec, timescale $\simeq$ 3 days) and the
thermal pulse 
(peak $L_{\nu} / L_{\nu,\odot} \simeq$ 10$^9$, flux $\Phi_{\nu, {\rm TP}} \simeq$ 1.7$\times$10$^7$ (10 pc/$d$)$^{2}$ cm$^{-2}$ s$^{-1}$,
timescale $\simeq$ 0.1 yr) phases of
evolution in low mass stars as potential probes for stellar neutrino
astronomy.  
We also delineate the contribution of neutrinos from
nuclear reactions and thermal processes to the total neutrino loss
along the stellar tracks in a neutrino Hertzsprung-Russell diagram.
We find, broadly but with exceptions, that neutrinos from nuclear
reactions dominate whenever hydrogen and helium burn, and that
neutrinos from thermal processes dominate otherwise.

\end{abstract}

% UAT concepts

\section{Introduction}\label{s.intro}

Stars radiate energy by releasing photons from the stellar surface and 
neutrinos from the stellar interior.  In the interior, weak
reactions produce electron neutrinos by thermal processes, electron and positron
captures on nuclei, and nuclear decays.  
Neutrinos interact feebly with baryonic matter, with typical cross sections 
of $\simeq$~10$^{-44}$~cm$^2$ as opposed to typical photon
cross sections of $\simeq$~10$^{-24}$~cm$^2$, escaping
from the star unhindered in circumstances where photons are trapped. 

Neutrino losses play key roles 
on the main-sequence in the case of the Sun 
\citep{bahcall_1992_aa,bahcall_2005_aa,haxton_2013_aa},
during the helium flash in red giants
\citep{ramadurai_1976_aa, sweigart_1978_aa,raffelt_1992_aa,catelan_1996_aa},
in the conversion of $^{14}$N to $^{22}$Ne during core helium burning
\citep{serenelli_2005_aa},
for the cooling of white dwarfs
\citep{van-horn_1971_aa,kawaler_1986_aa,fontaine_2001_aa,althaus_2010_aa,bischoff-kim_2018_aa},
during core carbon burning
\citep{ramadurai_1984_aa,aufderheide_1993_aa,meakin_2007_ab,cristini_2017_aa,cristini_2019_aa},
for pre-supernova stars
\citep{odrzywolek_2009_aa,kutschera_2009_aa,patton_2017_aa,patton_2017_ab},
for both core-collapse supernovae 
\citep[e.g.,][]{janka_2017_aa}
and electron-capture supernovae
\citep{ray_1984_aa,jones_2013_aa},
for the cooling of neutron stars
\citep{nomoto_1981_aa,potekhin_2015_aa},
during X-Ray bursts,
\citep{fujimoto_1987_aa,goodwin_2019_aa},
for accretion disks around black holes
\citep{birkl_2007_aa,fryer_2014_aa,uribe-suarez_2019_aa}
during neutron star mergers
\citep{albert_2017_aa,kyutoku_2018_aa},
and for nucleosynthesis
from the $\nu$-process  \citep{woosley_1990_aa}, $\nu\,p$ process \citep{mclaughlin_1995_aa, frohlich_2006_ab}, and r-process \citep[e.g.,][]{kajino_2019_aa}.

Neutrino production from thermal processes mainly depends on the ambient thermodynamic conditions
\citep{fowler_1964_aa, beaudet_1967_aa,schinder_1987_aa,itoh_1996_aa}. 
Neutrino production from electron/positron captures and nuclear decays have a 
stronger dependence on the isotopic composition
\citep{fuller_1980_aa,fuller_1982_aa,fuller_1982_ab,fuller_1985_aa, langanke_2000_aa,langanke_2014_aa,misch_2018_aa}, 
and thus on the network of nuclear reactions that take place in the stellar interior.  
These two classes of neutrino production thus carry complementary
information about the interior of stars \citep{patton_2017_aa,patton_2017_ab}.

Neutrino astronomy has been limited, so far, 
to the Sun 
\citep{borexino-collaboration_2018_aa},
supernova 1987A
\citep{hirata_1987_aa,hirata_1988_aa,bionta_1987_aa,alekseev_1987_aa},
and the blazar TXS 0506+056
\citep{icecube-collaboration_2018_aa,icecube-collaboration_2018_ab}.
However, the 
{\it Super-Kamiokande with Gadolinium} \citep{simpson_2019_aa},
{\it Jiangmen Underground Neutrino Observatory} \citep{li_2014_aa,brugiere_2017_aa},
and  {\it XENON} \citep{newstead_2019_aa} experiments
usher in a new generation of multi-purpose neutrino detectors 
designed to open new avenues for potentially observing currently undetected neutrinos.

This article is novel in exploring the evolution of stellar models in a
neutrino Hertzsprung-Russell (HR) diagram, where the traditional photon luminosity 
is replaced with the neutrino luminosity. This exploration provides targets for 
current, forthcoming, and future generations of neutrino detectors as well as
providing estimates of the stellar neutrino background signal. 
In Section~\ref{s.models} we describe the input physics and solar normalization of the stellar models.
In Section~\ref{s.nuhrd} we present our main results, and in 
Section~\ref{s.summary} we discuss and summarize our results.

\section{Stellar Models}\label{s.models}

\subsection{Input Physics}\label{s.input_physics}

We model the evolution of stars with initial masses M\,=\,1, 2, 3,
15, 25, 30, 35, and 40  \Msun\ from the pre-main sequence (PMS) to a
white dwarf (WD) for the lower masses, or the onset of core-collapse 
for the higher masses. These masses are chosen to delineate features in 
a forthcoming neutrino HR diagram. We use \MESA\ revision r12115 to
construct our stellar models \citep{paxton_2011_aa,paxton_2013_aa,paxton_2015_aa,paxton_2018_aa,paxton_2019_aa}.
Each star is modeled as a single, non-rotating, mass losing, solar metallicity object. 
The files to reproduce our work are publicly available at\dataset[http://doi.org/10.5281/zenodo.3634068]{http://doi.org/10.5281/zenodo.3634068}.

We use the built-in \MESA\ nuclear reaction network \texttt{mesa\_49} for low mass stars and 
\texttt{mesa\_204} for high mass stars. Relatively large nuclear networks are required to fully capture the 
energy generation rate, and thus the neutrino luminosity from $\beta$-processes, 
in neutron-rich compositions. The current defaults for nuclear reaction rates are
described in Appendix A.2 of \citet{paxton_2019_aa}.  Rates are taken from
a combination of NACRE \citep{angulo_1999_aa} and the Joint Institute for
Nuclear Astrophysics REACLIB library (default version, dated
2017-10-20) \citep{Cyburt_2010_ab}.  The \MESA\ screening corrections are
from \citet{chugunov_2007_aa}, which includes a physical parameterization
for the intermediate screening regime and reduces to the familiar weak
\citep{dewitt_1973_aa, graboske_1973_aa} and strong \citep{alastuey_1978_aa,itoh_1979_aa} 
limits at small and large values of the plasma coupling parameter.
All the weak reaction rates are based (in order of precedence) on the
tabulations of \citet{langanke_2000_aa},
\citet{oda_1994_aa}, and 
\citet{fuller_1985_aa}.

The three most dominant thermal neutrino processes are 
plasmon decay ($\gamma_{\rm plasmon} \rightarrow \nu_e + \bar{\nu}_e$), 
photoneutrino production ($e^- + \gamma \rightarrow e^- + \nu_e + \bar{\nu}_e$), 
and pair annihilation ($e^- + e^+ \rightarrow \nu_e + \bar{\nu}_e$).
The bremsstrahlung ($e^- + {\rm ^AZ} \rightarrow e^- + {\rm ^AZ} + \nu_e + \bar{\nu}_e$) 
and recombination ($e^-_{\rm continuum} \rightarrow e^-_{\rm bound} + \nu_e + \bar{\nu}_e$)
channels play smaller roles.
The total emissivities of all these processes, over a range of temperatures and densities, are discussed in
\citet{itoh_1989_aa,itoh_1992_aa,itoh_1996_aa,itoh_1996_ab}
and implemented in the \MESA\ thermal neutrino loss module.
Differential rates and emissivities of selected thermal neutrino processes are discussed in
\cite{ratkovic_2003_aa,dutta_2004_aa,misiaszek_2006_aa,odrzywoek_2007_aa,kato_2015_aa,patton_2017_aa,patton_2017_ab}.

The models approximate convection using the recipes described 
in \citet{paxton_2019_aa, paxton_2018_aa}.
The adopted values of the mixing-length parameter, $\alpha$ and overshooting parameter $f_{ov}$,
as well as the initial hydrogen fraction X, helium fraction Y, and metallicity Z
and are determined from our calibrated Solar model.

\subsection{Solar Neutrino Luminosity Normalization}\label{s.solar_norm}

We perform a Solar model calibration to reproduce the present day neutrino flux \citep{villante_2014_aa}.
We iterate on differences between the final model at $t_{\sun}$\,=\,4.568~Gyr \citep{2010NatGe...3..637B}
and the 
solar radius, 
$R_{\sun}= 6.9566 \times 10^{10}$ cm, 
solar luminosity,
$L_{\gamma,\sun}= 3.828 \times 10^{33}$ erg s$^{-1}$ \citep{2016AJ....152...41P}, 
and surface heavy element abundance Z/X. 
We use the built-in \MESA\ simplex module to iteratively vary 
the mixing-length parameter, $\alpha$, and the initial composition X, Y, and Z,
including the effects of element diffusion \citep{1994ApJ...421..828T,paxton_2018_aa}.
This calibration is performed for two estimates of the heavy element abundance at
the surface of the Sun, Z/X = 0.0181 \citep{asplund_2009_aa} and
Z/X = 0.0229 \citep{1998SSRv...85..161G}. We adopt a small amount of exponential convective overshooting \citep{2000A&A...360..952H} by choosing
$f_{ov}$ = 0.016 as used in the MIST isochrones \citep{2016ApJ...823..102C}. 
Separate implementations of convective overshooting at the base of the solar convection zone 
can be found in  \citet{2011MNRAS.414.1158C} and  \citet{2019ApJ...881..103Z}.
Our calibrated solar models do not include the structural
effects of rotational deformation or the effects of rotational mixing. 
Calibrated parameters are listed in Table \ref{tab:table1}.
We use the abbreviations
AGSS09 = \citet{asplund_2009_aa} photospheric abundances mixture and 
GS98 = \citet{1998SSRv...85..161G} meteoric abundance mixture in all Tables. 
The AGSS09 solar model is calculated using OPAL opacities \citep{1996ApJ...464..943I}, 
and the GS98 solar model is calculated using the Opacity Project (OP) opacities \citep{2005MNRAS.360..458B}. 
See \citet{2017ApJ...835..202V} for updated approaches toward standard solar models.

\begin{deluxetable}{cllc}[!htb]
  \tablenum{1}
  \tablecolumns{3}
  \tablewidth{\columnwidth}
  \tablecaption{Solar Calibration Parameters \label{tab:table1}}
  \tablehead{
    \colhead{Component} & \colhead{AGSS09} & \colhead{GS98}  }
  \startdata
      X\textsubscript{o}               & $0.7200$                &   0.7108                \\
      Y\textsubscript{o}               & $0.2654$                &  0.2710                          \\
      $\alpha_{mlt}$                   & $2.120$                &  2.155                      \\
      (Z/X)\textsubscript{surf}        & $0.0181$                      & 0.0229                     \\
      L$_{\nu,\odot}$/L$_{\gamma,\odot}$ & 0.02398             & 0.02422                                
  \enddata
\end{deluxetable}

\begin{deluxetable}{cllc}[!htb]
  \tablenum{2}
  \tablecolumns{4}
  \tablewidth{\columnwidth}
  \tablecaption{Properties of the Solar Calibrated Model \label{tab:table2}}
  \tablehead{
    \colhead{Component} & \colhead{AGSS09} & \colhead{GS98} & \colhead{Observed$^a$}}
  \startdata
      R\textsubscript{cz,b}/R$_{\odot}$   & $0.7256$          &          0.7178            & 0.713\,$\pm$\,0.001    \\
      Y\textsubscript{surf}                             & $0.2396$    &          0.2460            & 0.2485\,$\pm$\,0.0035  \\
  \enddata
    \tablenotetext{a}{The helioseismic derived 
radius at the bottom of the convective zone, R\textsubscript{cz,b}, 
and surface He mass fraction, Y\textsubscript{surf},
are from \citet{1997MNRAS.287..189B} and \citet{2004ApJ...606L..85B}.
}
\end{deluxetable}

\begin{deluxetable}{cllc}[!htb]
  \tablenum{3}
  \tablecolumns{4}
  \tablewidth{\columnwidth}
  \tablecaption{Solar Neutrino Fluxes \label{tab:table3}}
  \tablehead{
    \colhead{Component} & \colhead{AGSS09} & \colhead{GS98} & \colhead{Observed$^a$}}
  \startdata
      $\Phi$\textsubscript{pp} & 6.01 &  5.98 & $6.05(1^{+0.003}_{-0.011})$ \\
      $\Phi$\textsubscript{Be} & 4.71 &  4.95 & $4.82(1^{+0.05}_{-0.04})$   \\
      $\Phi$\textsubscript{B}  & 4.62 &  5.09 & $5.00(1\pm0.03)$            \\
      $\Phi$\textsubscript{N}  & 2.25 &  2.91 & $\leq6.7$                   \\
      $\Phi$\textsubscript{O}  & 1.67 &  2.21 & $\leq3.2$                   
  \enddata
    \tablenotetext{a}{Neutrino observations from the Borexino Collaboration \citep{2011PhRvL.107n1302B} as presented in \citet{haxton_2013_aa} and \citet{villante_2014_aa}.
The scales for neutrino fluxes $\Phi$ (in cm\textsuperscript{-2} s\textsuperscript{-1}) are: 
$10^{10}$ (pp); 
$10^{9}$  (Be);  
$10^{6}$  (B);  
$10^{8}$  (N); and  
$10^{8}$  (O).
}
\end{deluxetable}

% Place plot of solar sound speeds versus radius with estimated error
\begin{figure}[!htb]
\centering
\includegraphics[width=3.38in]{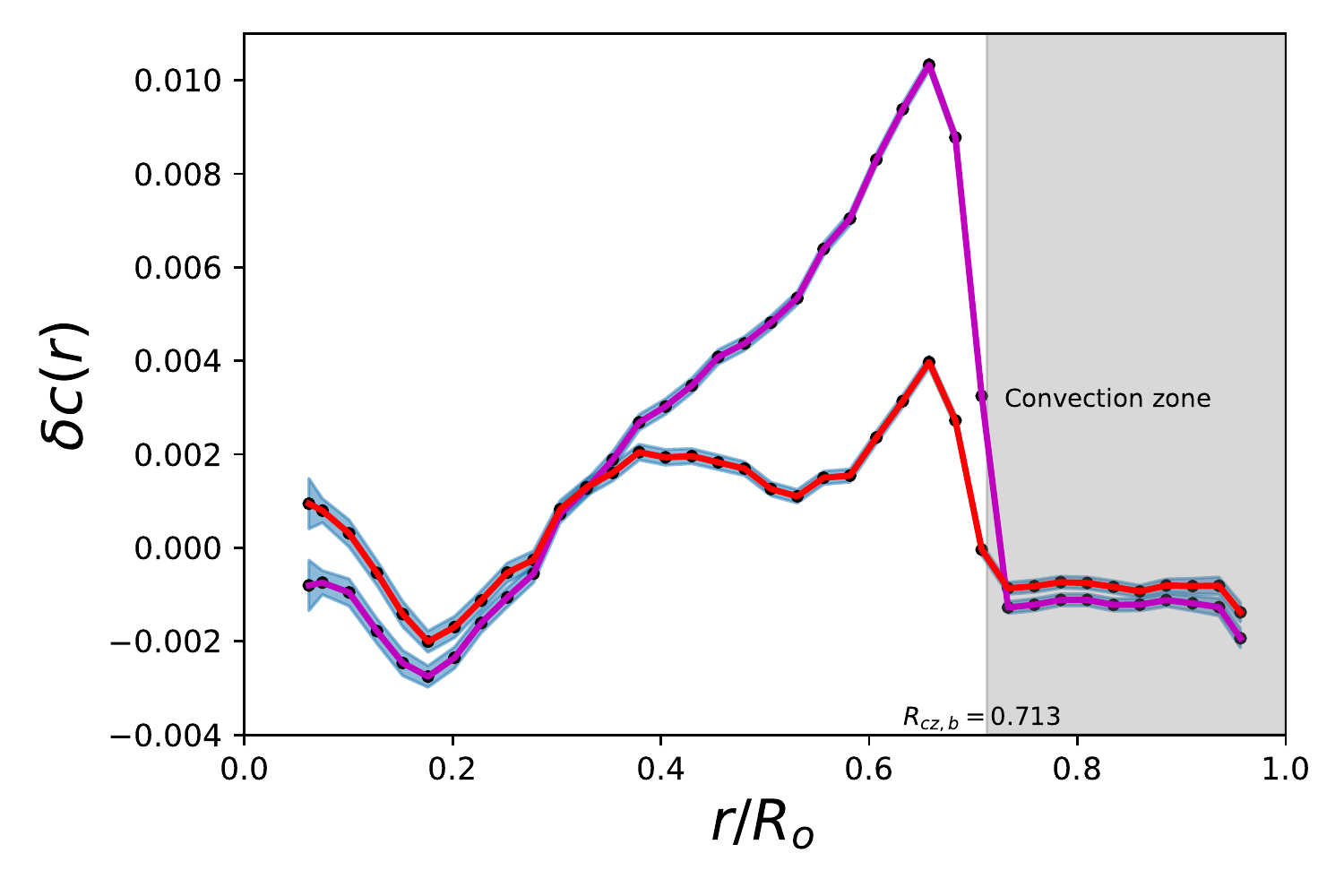} \\
\includegraphics[width=3.38in]{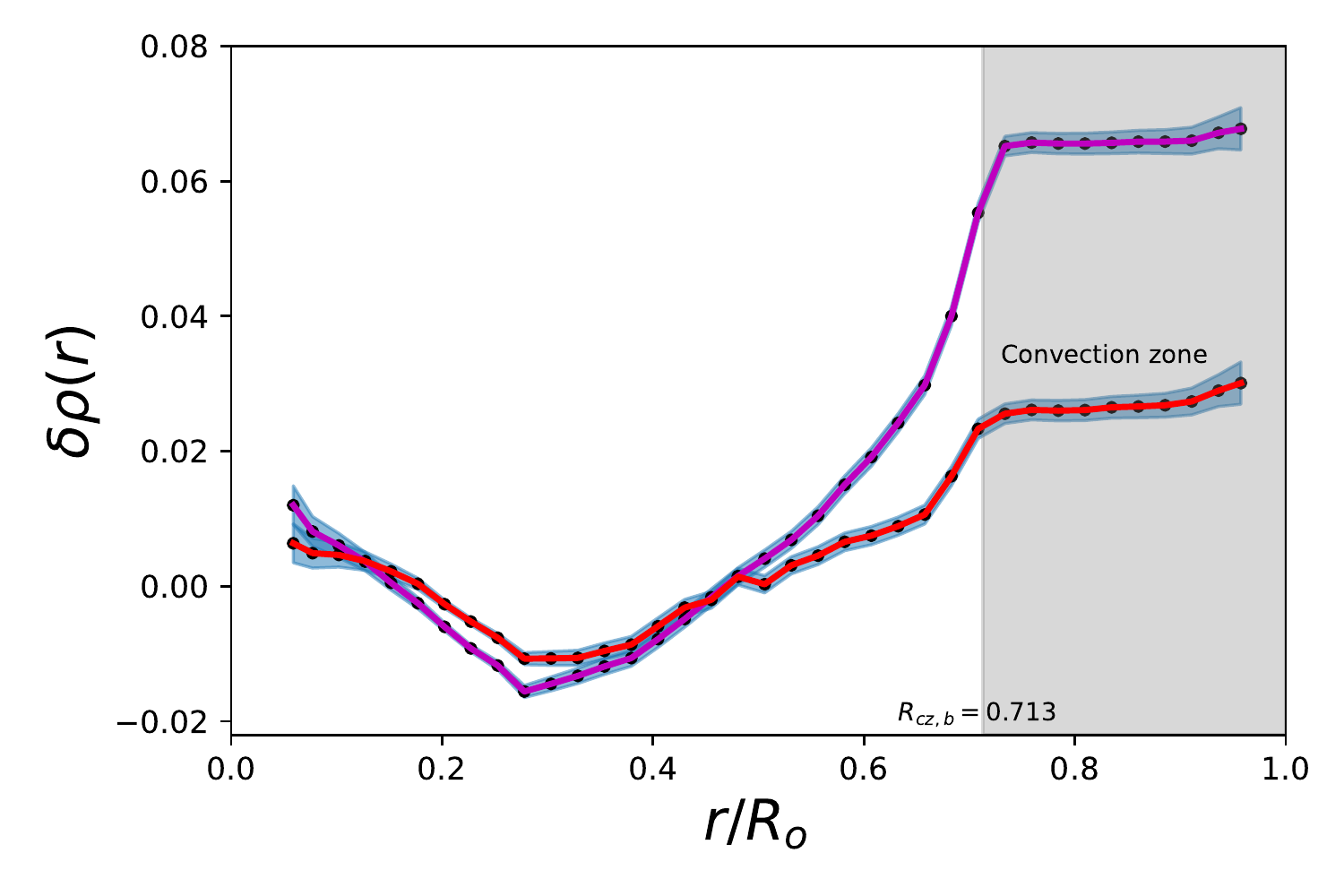}
\caption{
Fractional sound speed and density differences, 
$\delta c$ $=$ $(c_{\rm obs}$ - $c(r)) /c(r)$ and $\delta \rho$ $=$ $(\rho_{\rm obs}$ - $\rho(r)) /\rho(r)$, 
between the values predicted by the calibrated \MESA\ standard solar model, $c(r)$ and $\rho(r)$, 
and the values inferred from helioseismic data \citep{2009ApJ...699.1403B}, $c_{\rm obs}$ and $\rho_{\rm obs}$.
Black dots mark locations where $\delta c$ and $\delta \rho$ are evaluated.
Purple curves are for AGSS09 and red curves are for GS98. 
The gray band shows the convective region, with the radius at the base of the convection zone R\textsubscript{cz,b} marked.
The 3$\sigma$ uncertainties are shown as the blue bands.
}
\label{f.c_diff}
\end{figure}
% for this plot, include theoretical uncertainties in sounds speed
% also include uncertainties in helioseismic quanties from inversion
% use \delta{cr} as δci ≡ c_obs,i − c(ri ) /c(ri )

% use the png file for editing as the pdf rendering still takes too long
% but use the pdf upon submission - we want a high quality plot!
\begin{figure*}[!htb]
%\centering
\includegraphics[width=1.05\doublewide]{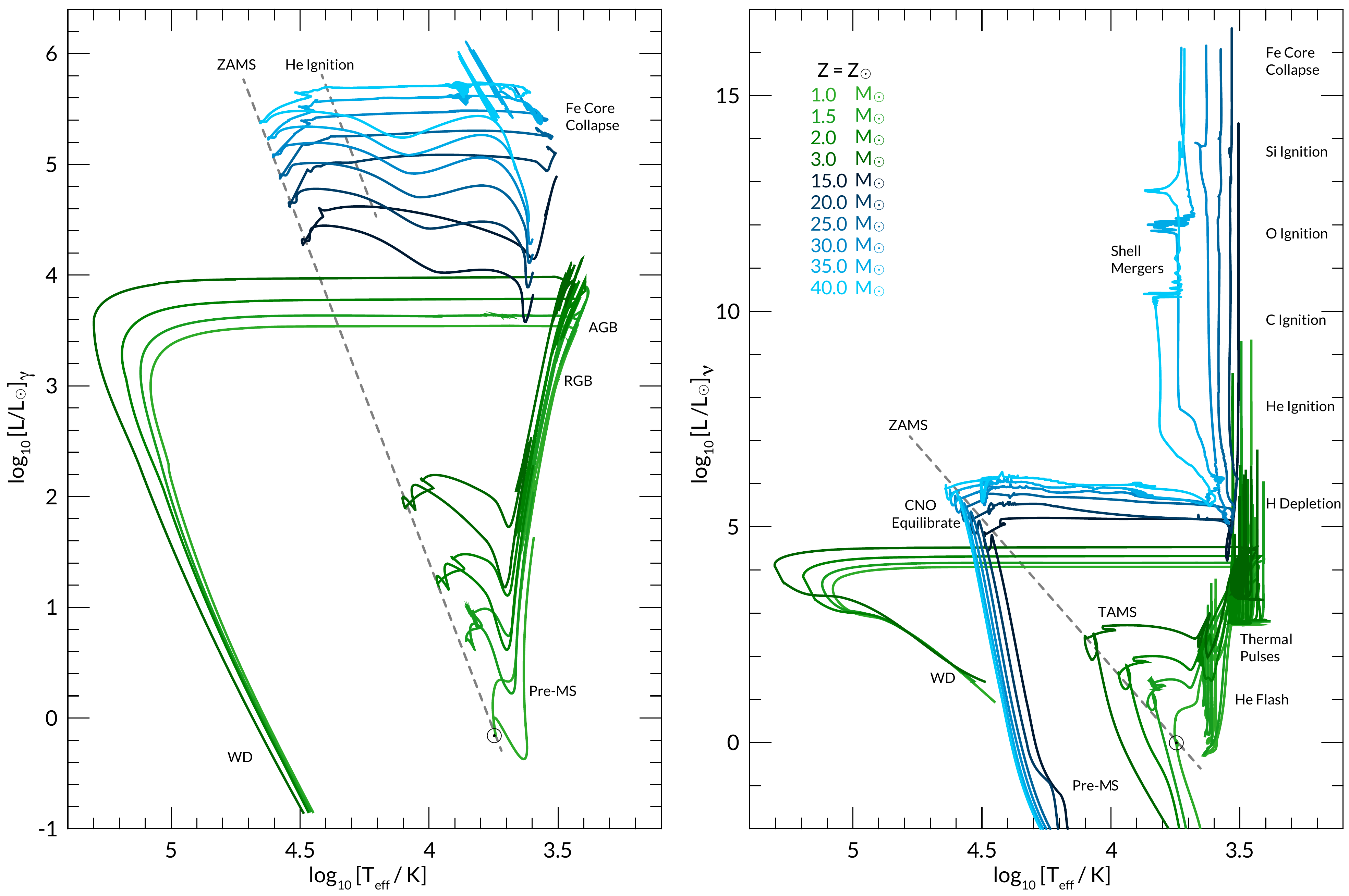}
\caption{
Stellar evolution tracks in a photon HR diagram (left) and a neutrino HR diagram (right).
Tracks for low mass stars are shades of green and those for high mass stars are shades of blue.
Luminosities are normalized by their respective current solar values,
$L_{\gamma,\odot}$ = 3.828 $\times$ 10$^{33}$ erg s$^{-1}$ \citep{2016AJ....152...41P}
and $L_{\nu,\odot}$ = 0.02398 $\cdot$ $L_{\gamma,\odot}$ = 9.1795 $\times$ 10$^{31}$ erg s$^{-1}$
(see Section \ref{s.solar_norm}), and key evolutionary phases are labeled. 
}
\label{f.hr_nu_hr}
\end{figure*}

Figure \ref{f.c_diff} shows the fractional difference in sound speed, $\delta c$,  and density, $\delta \rho$,  
between our calibrated solar models and the inferred helioseismic values, see \citet{2009ApJ...699.1403B}. 
Calculated values for helioseimic quantities are shown in Table \ref{tab:table2}.
Disagreements arise from differences in the 
solar abundance profiles,
equation of state,
opacities,
model atmospheres, 
treatment of convection, 
and the absence of rotational mixing. The task of correcting these disagreements is the subject of ongoing 
research, see \citet{2014dapb.book..245B} and \citet{2016EPJA...52...78S}.
Nonetheless, our seismic results appear similar to those in \citet{villante_2014_aa} and \citet{asplund_2009_aa}.

Neutrinos are produced during H-burning on the main-sequence (MS)
from the proton-proton (pp) chain reactions
p(p,e$^+$$\nu_e$)$^2$H, 
p(e$^-$p,$\nu_e$)$^2$H, 
$^3$He(p,e$^+$$\nu_e$)$^4$He, 
$^7$Be(e$^-$,$\nu_e$)$^7$Li, 
$^8$B(,e$^+$$\nu_e$)$^8$Be,
and the CNO cycle reactions
$^{13}$N(,e$^+$$\nu_e$)$^{13}$C, 
$^{13}$N(e$^{-}$,$\nu_e$)$^{13}$C,
$^{15}$O(,e$^+$$\nu_e$)$^{15}$N, 
$^{15}$O(e$^{-}$,$\nu_e$)$^{15}$N,
$^{17}$F(,e$^+$$\nu_e$)$^{17}$O, 
$^{17}$F(e$^{-}$,$\nu_e$)$^{17}$O,
$^{18}$F(,e$^+$$\nu_e$)$^{18}$O,
where electron capture reactions on CNO nuclei are included \citep{stonehill_2004_aa}.
Higher temperatures can trigger the production of nuclear reaction neutrinos from the 
H-burning hot CNO, Ne-Na, and Mg-Al cycles. 

The neutrino flux in the solar interior is strongly dependent on the
core temperature \citep[see][]{1996PhRvD..53.4202B}. Standard solar models
that accurately predict temperatures near the solar core should also 
generate comparable neutrino fluxes to solar neutrino data. 
Neutrino fluxes are calculated from each solar model and compared to observations in Table \ref{tab:table3}.
Our predicted neutrino fluxes are similar to \citet{villante_2014_aa} and  \citet{haxton_2013_aa}.  
We adopt the AGSS09 \MESA\  model, calculated using OPAL opacities, as the standard in this article.
Specifically, we use $L_{\nu,\odot}$ = 0.02398 $\cdot$ $L_{\gamma,\odot}$ = 9.1795 $\times$ 10$^{31}$ erg s$^{-1}$
as the normalization for the neutrino HR diagram.

\section{Evolution in A Neutrino HR Diagram}\label{s.nuhrd}

Stars are powered mainly by fusion reactions throughout their life,
but weak reactions play a key role in determining their structure, energy
budget, and nucleosynthesis. 
A fundamental aspect of weak reactions for stellar evolution is that
they facilitate hydrogen fusion into helium \citep[for Universes and stellar evolution without the weak force see][]{grohs_2018_aa}.
They affect the interior structure
because the pressure is mostly due to free electrons and in some cases (e.g., electron capture supernovae) weak
reactions change the number of free electrons. Neutrino
losses modify the energy budget, and dominate for C-burning and beyond. 
Finally, they affect the nucleosynthesis because the production 
of most nuclei is sensitive to the electron to baryon ratio.

Figure \ref{f.hr_nu_hr} shows the stellar evolution tracks of the
models considered in a photon and neutrino HR diagram.  A photon HR
diagram uses two surface properties, the effective temperature $T_{\rm eff}$ 
and photon luminosity $L_{\gamma}$.  A neutrino HR diagram uses the
 $T_{\rm eff}$ surface property and an interior property, the
neutrino luminosity $L_{\nu}$.  We next discuss the key phases of evolution
that are labeled in the neutrino HR diagram.

Each pre-main sequence (PMS) model begins with a uniform composition and central
temperature that is low enough that nuclear burning is
inconsequential.  The central temperature and density then increase as the
stellar model undergoes gravitational contraction.  The initial CNO
abundances for solar metallicity stars is not equal to the CNO
abundances when the CNO cycle is operating in equilibrium.  Nuclear
reactions replace gravitational contraction as the major source of
$L_{\gamma}$ and $L_{\nu}$ by burning the $^{12}$C abundance 
to a value that is commensurate with CNO equilibrium
values \citep{iben_1965_aa}.

The reactions $^{12}$C($p$,$\gamma$)$^{13}$N(,$e^{+} \nu_e$)$^{13}$C($p$,$\gamma$)$^{14}$N
can occur at lower temperatures than when the full CNO cycle competes
with the pp-chain.  They produce a nuclear energy 
$E_{\rm nuc}\,\simeq\,N_A Q \rho X_{c} / {\rm A}_c$, 
where $N_A$ is the Avogadro number,
$\rho$ is the mass density, $X_c$ is the mass fraction of $^{12}$C,
A$_c$ is the number of nucleons in $^{12}$C, and $Q \simeq$~11 MeV
is the nuclear binding energy release.  The thermal energy 
is $E_{\rm th} \simeq 3/2 N_A \rho k_B T$, where $k_B$ is the
Boltzmann constant and $T$ is the temperature.  The ratio 
at solar metallicity and $T$~=~10$^7$~K is 
$E_{\rm  nuc} / E_{\rm th} \simeq Q X_{c} /(18 k_B T) \simeq$\,1.5 
\citep{bildsten_2019_aa}.  
That is, the star can delay gravitational contraction for about one
Kelvin-Helmholtz by reducing $^{12}$C. This transition from
the PMS to the zero-age main sequence (ZAMS) is visible in the neutrino
HR diagram of Figure \ref{f.hr_nu_hr} as the loop prior to landing on
the ZAMS.

For all of the models considered, core H-burning powers $L_{\nu}$ on the 
ZAMS by the weak reactions given in Section \ref{s.solar_norm}.  
As H in the core depletes, all the models enter the terminal-age main sequence 
(labeled TAMS in Figure \ref{f.hr_nu_hr})
and continue to evolve toward cooler $T_{\rm eff}$.
Further evolution is now divided into low mass stars (Section \ref{s.lowmass})
and high mass stars (Section \ref{s.highmass}).

\subsection{Low Mass Stars}\label{s.lowmass}

Low mass stars (M $\lesssim$ 8 \Msun) ascend the red giant branch 
(labeled RGB in Figure \ref{f.hr_nu_hr}) 
as they evolve to cooler $T_{\rm eff}$ in the photon HR diagram, and 
evolve at approximately constant $L_{\nu}$ from shell H-burning 
in the neutrino HR diagram. 

As stars evolve, the ashes of nuclear burning usually have
a heavier mean atomic number and lie interior to the unburned fuel. 
For example, the He core is interior to the H-burning shell, and 
the CO core is interior to the He-burning shell.
One class of exceptions occurs when a combination of electron degeneracy
and thermal neutrino losses lead to cooler temperatures in the
central regions and the fuel ignites off-center. Examples include 
He ignition in $M_{\rm ZAMS} \lesssim 2 M_{\odot}$ stars (i.e., the ``helium flash'')
and C ignition in Super-AGB stars.
Fuels that ignite off-center develop convection behind the nuclear burning 
(towards the surface of the star) and propagate towards the center.  
These convectively bounded flames have relatively slow speeds 
{\citep{timmes_1994_ab,garcia-berro_1997_aa,schwab_2020_aa}, 
due to the propagation being driven by 
thermal conduction under semi-degenerate conditions.

Helium ignition occurs at the tip of the RGB in the photon HR diagram
and in the lower-right in the neutrino HR diagram.  
The slowest step in the H-burning CNO cycle is the proton capture onto $^{14}$N. 
This results in all the CNO catalysts piling up into $^{14}$N when core H-burning
is complete. During He-burning all of the $^{14}$N is converted in $^{22}$Ne
by the reaction sequence
$^{14}$N($\alpha$,$\gamma$)$^{18}$F(,$e^{+}\nu_e$)$^{18}$O($\alpha$,$\gamma$)$^{22}$Ne.
It is the weak reaction in this sequence that powers $L_{\nu}$ throughout this 
phase of evolution \citep[e.g.,][]{serenelli_2005_aa}.

The He core flash phase, which occurs in $M_{\rm ZAMS} \lesssim 2\,M_{\odot}$ stars,
is characterized by a series of subflashes
that propagate toward the stellar center
\citep{thomas_1967_aa,serenelli_2005_ab,bildsten_2012_ab,gautschy_2012_aa,serenelli_2017_aa}.
For example, the 1 \Msun\ model in Figure \ref{f.hr_nu_hr} undergoes
five subflashes with the first subflash occurring at
$\simeq$~0.18~\Msun and reaching $L_{\nu}$ $\simeq$ 10$^4$ $L_{\nu,\odot}$.
The number of subflashes decreases as the stellar mass increases, 
and the initial flash takes place closer to the stellar center.  
These subflashes, with their dependence on the stellar mass, are
visible in the neutrino HR diagram of Figure \ref{f.hr_nu_hr} as the
spikes in the region labeled ``He Flash''.
After the He core flash phase, which burns very little helium, core He-burning then proceeds quiescently
\citep[e.g.,][]{deboer_2017_aa} to produce an electron degenerate CO core.

Helium ignition in $M_{\rm ZAMS} \gtrsim 2 M_{\odot}$} stars
occurs under non-degenerate conditions, without flashes or subflashes,
and leads to a different, smoother, signature in the production of neutrinos 
from $^{18}$F decay.
For the 2\,M$_{\odot}$ model, from Figure \ref{f.hr_nu_hr}, 
$L_{\nu}$=0.8\,$L_{\gamma,\odot}$ and $L_{\gamma}$=16\,$L_{\gamma,\odot}$ on the MS,
$L_{\nu}$=120\,$L_{\gamma,\odot}$ and $L_{\gamma}$=1750\,$L_{\gamma,\odot}$ at He-ignition (tip of the RGB),
$L_{\nu}$=5.1\,$L_{\gamma,\odot}$ and $L_{\gamma}$=110\,$L_{\gamma,\odot}$ at core He-depletion (mass fraction of $^4$He less than 0.001),
$L_{\nu}$=420\,$L_{\gamma,\odot}$ and $L_{\gamma}$=6100\,$L_{\gamma,\odot}$ after the thermal pulses when the envelope mass is 0.01\,M$_{\odot}$,
$L_{\nu}$=1.8\,$L_{\gamma,\odot}$ when $L_{\gamma}$=1.0\,$L_{\gamma,\odot}$ on the WD cooling track.

\begin{figure*}[!htb]
\centering
\includegraphics[width=0.53\textwidth]{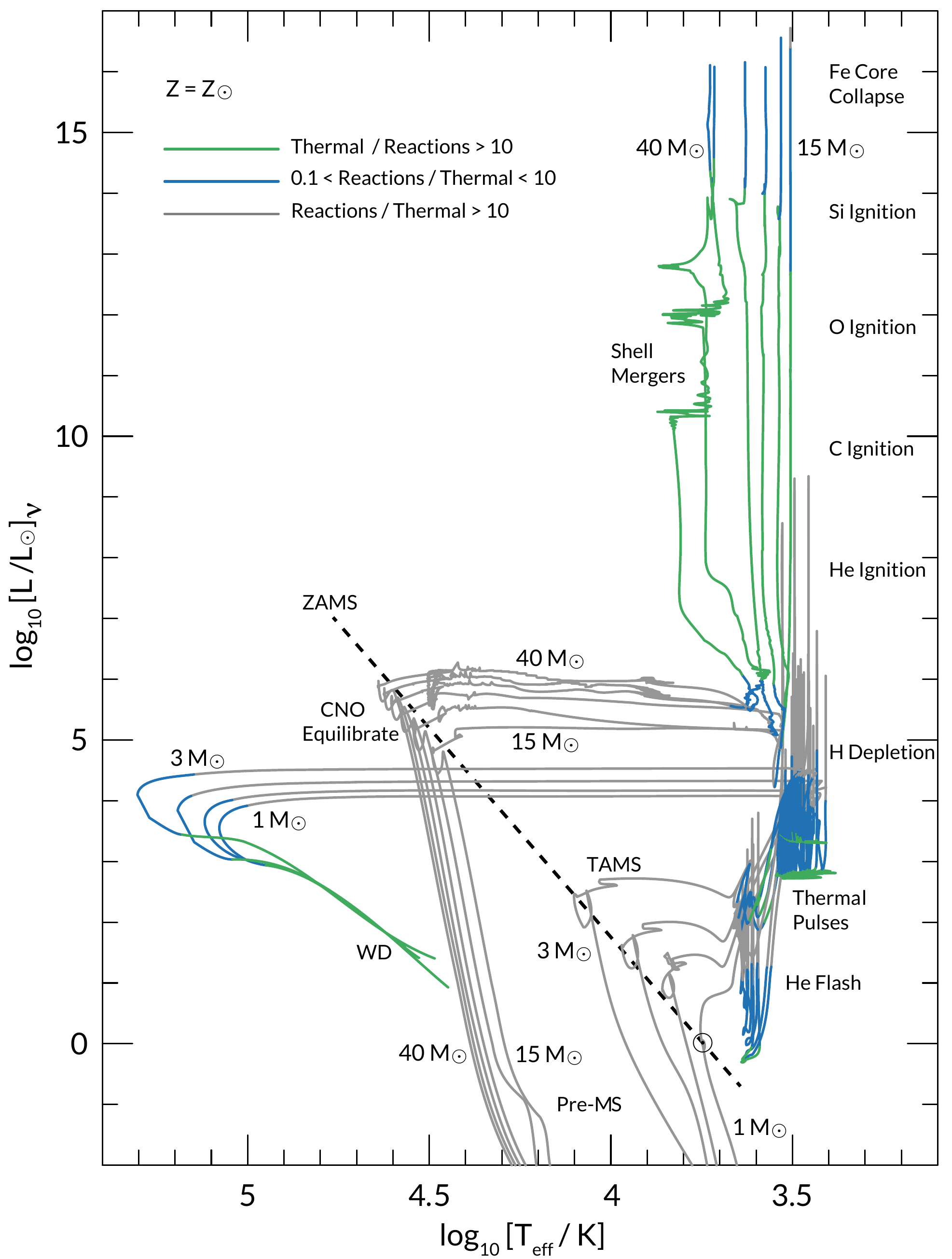}
\caption{
Ratio of the nuclear reaction neutrino luminosity to the thermal neutrino luminosity 
plotted along stellar evolution tracks in a neutrino HR diagram.
Gray curves indicate where nuclear reaction neutrinos dominate,
green curves where thermal neutrinos dominate, and
blue curves where the reaction and thermal neutrino luminosities 
are within a factor of 10. 
The neutrino luminosity is normalized by the current solar value
$L_{\nu,\odot}$ = 0.02398 $\cdot$ $L_{\gamma,\odot}$ = 9.1795 $\times$ 10$^{31}$ erg s$^{-1}$
(see Section \ref{s.solar_norm}).
}
\label{f.hr_nu_reac_therm}
\end{figure*}

Asymptotic giant branch (AGB) stars are the final stage of evolution
driven by nuclear burning. This phase is characterized by H and He burning in geometrically thin shells on
top of the CO core \citep{herwig_2005_aa}. For the more massive super-AGB stars 
a ONeMg core is produced from a convectively bounded carbon flame 
that propagates toward the center \citep{becker_1979_aa,becker_1980_aa,garcia-berro_1997_aa,
siess_2007_aa,denissenkov_2015_aa,farmer_2015_aa,lecoanet_2016_ab}.

A thin He shell grows as material from the adjacent H-burning shell is processed,
causing the He shell to increase in temperature and pressure. Once the
mass in the He shell reaches a critical value, He ignition causes
a thermal pulse.  For example, the 3 \Msun\ model goes through a
series of six thermal pulses, with an interpulse period of $\simeq$
10$^5$ yr.  
The number of thermal pulses a model undergoes is poorly determined 
as the number is sensitive to the mass resolution,
the stellar mass loss rate, and the treatment of convective boundaries.
These thermal pulses are visible in the neutrino HR
diagram of Figure \ref{f.hr_nu_hr} as the spikes in the region 
labelled ``Thermal Pulses''.

The stellar models leave the thermal pulse phase when the envelope mass above the
still active H and He burning shells is reduced to $\simeq$ 0.01 \Msun\ by stellar winds. All the low mass models 
then evolve toward larger $T_{\rm eff}$ at nearly
constant $L_{\nu}$ and $L_{\gamma}$.  Nuclear burning extinguishes as
the post-AGB model enters the WD cooling track.  Plasmon neutrino emission then
dominates the energy loss budget for average-mass CO WDs with 
$T_{\rm eff}$~$\gtrsim$~25,000~K \citep{vila_1966_aa,kutter_1969_aa,bischoff-kim_2018_aa}.  
As the WD continues to cool, photons
leaving the surface begin to dominate the cooling as the electrons
transition to a strongly degenerate plasma \citep{van-horn_1971_aa,corsico_2019_aa}. The low mass models in 
Figure \ref{f.hr_nu_hr} are arbitrarily chosen to terminate when the WD reaches $L_{\gamma}$ = 0.1 $L_{\gamma,\odot}$.
With $T_{\rm eff}$~$\simeq$~30,000~K at this arbitrary termination point, the WD models are still dominated
by thermal neutrino cooling, $L_{\nu}/L_{\gamma} \simeq 3$ \citep{winget_2004_aa}. For calculating the integrated neutrino background from stellar sources, especially
if WDs are abundant, these models should be further evolved to $T_{\rm eff}$~$\lesssim$~12,000~K
to drive $L_{\nu}/L_{\gamma} \le 10^{-5}$ \citep[e.g., Figure 5 in][]{timmes_2018_ab}.

\begin{figure*}[!htb]
\centering
\includegraphics[width=0.95\textwidth]{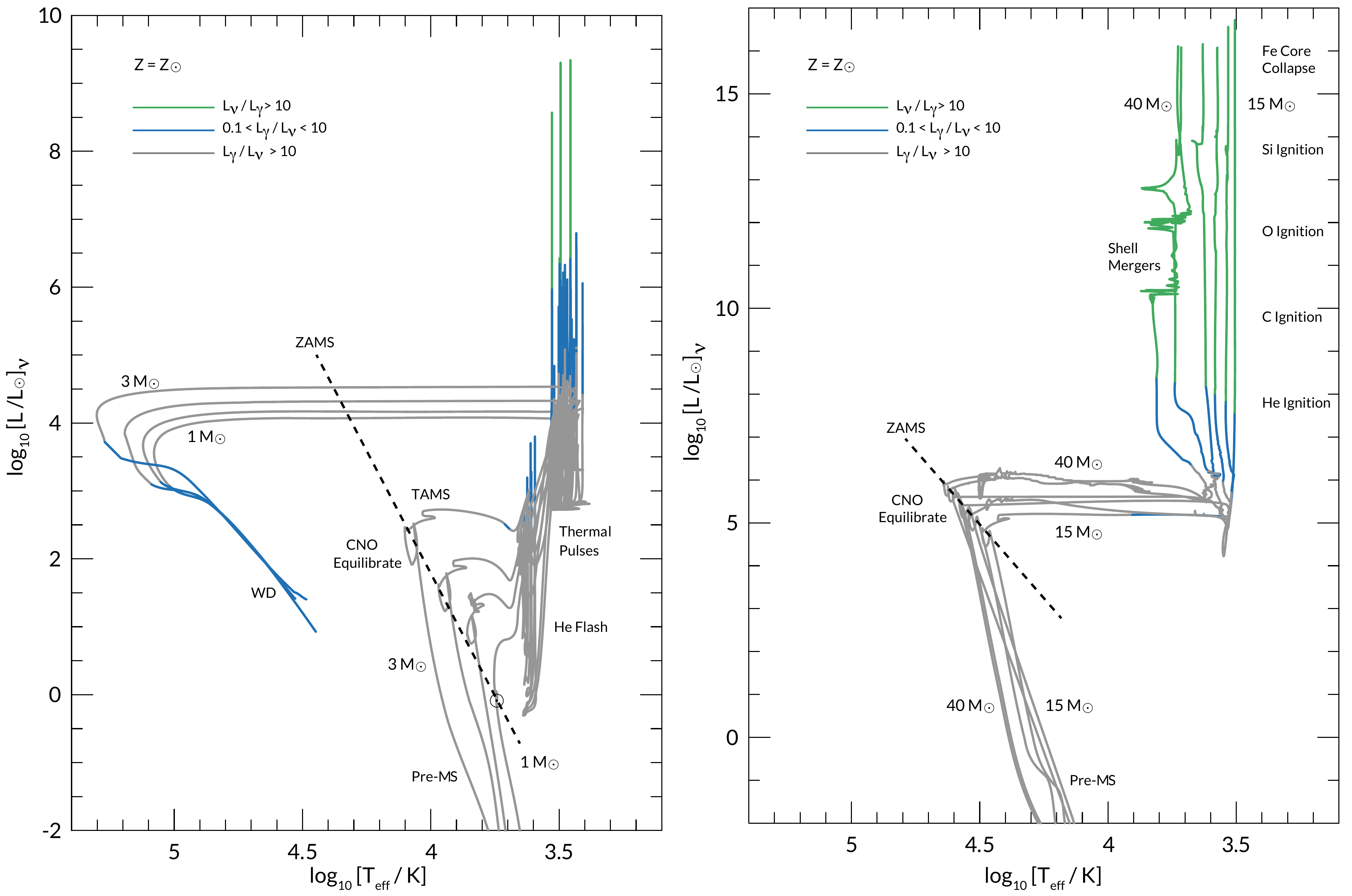}
\caption{
Ratio of the photon luminosity to the neutrino luminosity in a neutrino HR diagram.
Low mass stars are on the left, high mass stars on the right.
Gray curves indicate where photons dominate,
green curves where neutrinos dominate, and
blue curves where the photon and neutrino luminosities 
are within a factor of 10. 
The neutrino luminosity is normalized by the current solar value
$L_{\nu,\odot}$ = 0.02398 $\cdot$ $L_{\gamma,\odot}$ = 9.1795 $\times$ 10$^{31}$ erg s$^{-1}$
(see Section \ref{s.solar_norm}).
}
\label{f.hr_nu_gamma_nu}
\end{figure*}

\subsection{High Mass Stars}\label{s.highmass}

High mass stars (M $\gtrsim$ 8 \Msun) in Figure \ref{f.hr_nu_hr}
evolve at nearly constant $L_{\nu}$ and $L_{\gamma}$ as hydrogen depletes 
in the core and the models evolve to cooler $T_{\rm eff}$.
Free streaming neutrinos from thermal processes, primarily pair annihilation,
dominate a star's energy loss budget from the core C-burning phase
to core Si depletion. 
For the 30\,M$_{\odot}$ model, from Figure \ref{f.hr_nu_hr}, 
$L_{\nu}$=8.2$\times$10$^{3}$\,$L_{\gamma,\odot}$ and $L_{\gamma}$=1.2$\times$10$^{5}$\,$L_{\gamma,\odot}$ on the MS,
$L_{\nu}$=1.3$\times$10$^{4}$\,$L_{\gamma,\odot}$ and $L_{\gamma}$=3.0$\times$10$^{5}$\,$L_{\gamma,\odot}$ at core He-ignition,
$L_{\nu}$=5.3$\times$10$^{3}$\,$L_{\gamma,\odot}$ and $L_{\gamma}$=2.7$\times$10$^{5}$\,$L_{\gamma,\odot}$ at core He-depletion 
(mass fraction of $^4$He less than 0.001),
$L_{\nu}$=3.2$\times$10$^{7}$\,$L_{\gamma,\odot}$ and $L_{\gamma}$=3.1$\times$10$^{5}$\,$L_{\gamma,\odot}$ at core C-ignition.
This dominance over photons as the primary
energy loss mechanism sets a rapid evolutionary timescale (years to hours) for
the advanced stages of nuclear fusion in presupernova stars
\citep{woosley_2002_aa}.  This rapid evolution is visible in the
neutrino HR diagram of Figure \ref{f.hr_nu_hr} as the nearly vertical curves
at approximately constant $T_{\rm eff}$.

Weak reactions that increase the electron to baryon ratio during C-burning include
$\beta$-processes involving $^{23}$Mg and $^{21,22}$Na.
The composition continues to become more neutron-rich 
during O-burning 
from $\beta$-processes on $^{30,33}$P, $^{33}$P, $^{35}$Cl, and $^{37}$Ar
Core Si-burning is the last exothermic burning stage and produces the Fe-peak nuclei.
Many isotopes in this stage of evolution undergo $\beta$-processes that continue to make the material more
neutron-rich \citep[see][]{heger_2001_aa,odrzywolek_2009_aa,patton_2017_ab}.

Dynamical large-scale mixing on nuclear burning timescales occurs during the 
late stages of evolution in massive stars. Stellar evolution models 
suggest that merging occurs between the C, Ne, O, and Si shells.  
These shell mergers are beginning to be explored with 3D hydrodynamic simulations
\citep[e.g.,][]{ritter_2018_aa}. The approximate location of these shell mergers is labeled
in the neutrino HR diagram. In addition, the energetics of nuclear burning
tightly couples to turbulent convection during O-burning and Si-burning.
This strong coupling must be modeled with 3D simulations 
\citep{meakin_2007_ab,couch_2015_aa,muller_2017_aa,fields_2020_aa}
to assess the fidelity of the convection approximations made by 1D models.

When the Fe core reaches its finite-temperature Chandrasekhar mass,
electron capture and photodisintegration drive collapse of the Fe
core, with the largest infall speeds usually occurring 
near the outer edge of the Fe core. The massive star models in Figure \ref{f.hr_nu_hr}
terminate when any mass coordinate within the Fe core exceeds an
inward velocity of 300 km sec$^{-1}$.

\subsection{Reaction and Thermal Neutrino Luminosities}\label{s.reac2therm}

Figure \ref{f.hr_nu_reac_therm} shows the ratio of nuclear reaction neutrinos to thermal
neutrinos along the stellar evolution tracks in the neutrino HR diagram.
Broadly, neutrinos from reactions dominate during H and He burning, 
and thermal neutrinos dominate for C-burning onwards. 
There are exceptions to this general scenario. One exception
is between the subflashes of the He flash for low mass stars, 
where thermal neutrinos become comparable or larger than neutrino losses from reactions.
Another exception are between thermal pulses on the AGB where
thermal neutrinos are again comparable or larger than nuclear reaction neutrinos.
Conversely, nuclear reaction neutrinos are comparable to, but less than, 
thermal neutrinos during the final phases of massive star evolution.

\subsection{Photon and Neutrino Luminosities}\label{s.gamma2nu}

Figure \ref{f.hr_nu_gamma_nu} shows the $L_{\gamma} / L_{\nu}$ ratio 
along the stellar evolution tracks in the neutrino HR diagram.
Photons dominate over most of star's lifetime \citep[e.g.,][]{barkat_1975_aa}, except in the advanced
stages of evolution, where neutrinos dominate on the early portions of the WD cooling tracks for low mass stars and
for carbon burning to the onset of core collapse for high mass stars.

\section{Discussion and Summary}\label{s.summary}

Using a \MESA\ solar calibrated model for the Sun's neutrino luminosity
as a normalization (Section \ref{s.solar_norm}), we have explored the
evolution of a select grid of stellar models from their pre-main sequence phase
to near their final fates in a neutrino HR diagram (Figure \ref{f.hr_nu_hr}). 
We also delineated the contributions from reaction and
thermal neutrinos during a model's evolution (Figure \ref{f.hr_nu_reac_therm}).  
This is the first time, to our knowledge, that such an exploration with a 
different messenger, neutrinos, has been presented in the literature.

Neutrino astronomy is a unique tool that can yield insights into
otherwise hidden aspects of stellar astrophysics \citep{bahcall_1989_aa,beacom_2010_aa}. 
However, the small cross section between neutrinos and baryonic matter, which allows
neutrinos to escape from the star in the first place, means it is
unlikely that near-future neutrino detectors will be able
to probe the neutrino luminosity tracks shown in Figure \ref{f.hr_nu_hr}. 

A possible exception is the evolution of a pre-supernova star on timescales
of a $\simeq$10 hr before Fe core-collapse. 
For a normal neutrino mass hierarchy, more than 200 events could be 
detected before core collapse for a 
15-30 $M_{\odot}$ star at $\simeq$ 200 pc (e.g., $\alpha$ Orionis, Betelgeuse),
and neutrino emission may be detectable
within $\simeq$ 600 pc with the improved sensitivity of Super-Kamiokande with Gadolinium
\citep{patton_2017_ab,simpson_2019_aa}.

Another possible exception is the detection of neutrinos from the He flash
and thermal pulses of low mass stars. Figures \ref{f.hr_nu_hr} and
\ref{f.hr_nu_reac_therm} suggest the He flash reaches peaks of
$L_{\nu} \simeq$ 10$^4$ $L_{\nu,\odot}$ and is driven by the
$^{18}$F(,$e^{+}\nu_e$)$^{18}$O reaction \citep{serenelli_2005_aa}.
The maximum energy of neutrinos emitted by this reaction is
$\simeq$~0.6~MeV and the average energy is $\simeq$~0.3~MeV.  
The neutrino flux is thus
$\Phi_{\nu, {\rm He \ flash}} \simeq$ 170 (10 pc/$d$)$^{2}$ cm$^{-2}$ s$^{-1}$
for a star located at a distance of $d$ parsec. The timescale of this peak emission is $\simeq$ 3 days,
depending chiefly on the initial ZAMS mass.
Figures \ref{f.hr_nu_hr} and \ref{f.hr_nu_reac_therm} also suggest
that the He-burning driven thermal pulses reach peaks of $L_{\nu}~\simeq$~10$^9$~$L_{\nu,\odot}$
from the same $^{18}$F(,$e^{+}\nu_e$)$^{18}$O reaction 
with an average energy of $\simeq$~0.3~MeV.
This gives a neutrino flux  of
$\Phi_{\nu, {\rm TP}} \simeq$ 1.7$\times$10$^7$ (10 pc/$d$)$^{2}$ cm$^{-2}$ s$^{-1}$
on timescales of $\simeq$ 0.1 yr, depending on the mass of the 
stellar envelope, uncertain mass loss rate, and pulse number.
Finally, integration of the neutrino luminosity
stellar evolution tracks may be useful
for refining estimates of the diffuse stellar neutrino background 
\citep{horiuchi_2009_aa,beacom_2010_aa}.

\acknowledgements

% people and places shout-outs

We thank the anonymous referee for improving this article, and 
Aaron Dotter, Thomas Steindl, and Josiah Schwab for discussions.
The \MESA\ project is supported by the National Science Foundation (NSF)
under the Software Infrastructure for Sustained Innovation program grants 
(ACI-1663684, ACI-1663688, ACI-1663696).
This research was also supported by 
the NSF under grant PHY-1430152 for the Physics Frontier Center ``Joint Institute
for Nuclear Astrophysics - Center for the Evolution of the Elements'' (JINA-CEE).
RF is supported by the Netherlands Organization for Scientific Research (NWO) 
through a top module 2 grant with project number 614.001.501 (PI de Mink).
FXT acknowledges stimulating discussions at Sky House.
This research made extensive use of the SAO/NASA Astrophysics Data System (ADS).

\software{
\MESA\ \citep[][\url{http://mesa.sourceforge.net}]{paxton_2011_aa,paxton_2013_aa,paxton_2015_aa,paxton_2018_aa,paxton_2019_aa},
\texttt{MESASDK} 20190830 \citep{mesasdk_linux,mesasdk_macos},
\texttt{matplotlib} \citep{hunter_2007_aa}, and
\texttt{NumPy} \citep{der_walt_2011_aa}.
         }

\bibliographystyle{aasjournal}
%\bibliography{paper}

\end{document}